\documentclass[12pt]{elsarticle}
\usepackage{refmerge}
\usepackage{epsfig}
\def\Ecm {\ensuremath{\rm E_{\rm c.m.}}}
\def\BR {\ensuremath{\mathcal B}}
\def\epem {\ensuremath{e^+ e^-}}
\def\pipi {\ensuremath{\pi^+\pi^-}}
\def\mevc {\ensuremath{\rm MeV/c}}
\def\mevcc {\ensuremath{\rm MeV/c^2}}
\def\piz {\ensuremath{\pi^0}}
\begin{document}
\date{\today}

\title{\bf{ \boldmath
STUDY OF THE PROCESS $e^+e^-\to \pi^+\pi^-\pi^0\eta$
IN THE C.M. ENERGY RANGE 1394--2005 MEV WITH THE CMD-3 DETECTOR
}}

\author[adr1,adr2]{R.R.~Akhmetshin}
\author[adr1,adr2]{A.N.~Amirkhanov}
\author[adr1,adr2]{A.V.~Anisenkov}
\author[adr1,adr2]{V.M.~Aulchenko}
\author[adr1]{V.Sh.~Banzarov}
\author[adr1]{N.S.~Bashtovoy}
\author[adr1,adr2]{D.E.~Berkaev}
\author[adr1,adr2]{A.E.~Bondar}
\author[adr1]{A.V.~Bragin}
\author[adr1,adr2]{S.I.~Eidelman}
\author[adr1,adr2]{D.A.~Epifanov}
\author[adr1,adr2,adr3]{L.B.~Epshteyn}
\author[adr1,adr2]{A.L.~Erofeev}
\author[adr1,adr2]{G.V.~Fedotovich}
\author[adr1,adr2]{S.E.~Gayazov}
\author[adr1,adr2]{A.A.~Grebenuk}
\author[adr1,adr2]{S.S.~Gribanov}
\author[adr1,adr2,adr3]{D.N.~Grigoriev}
\author[adr1,adr2]{F.V.~Ignatov}
\author[adr1,adr2]{V.L.~Ivanov}
\author[adr1]{S.V.~Karpov}
\author[adr1,adr2]{V.F.~Kazanin}
\author[adr1,adr2]{I.A.~Koop}
\author[adr1]{A.N.~Kirpotin}
\author[adr1,adr2]{A.A.~Korobov}
\author[adr1,adr3]{A.N.~Kozyrev}
\author[adr1,adr2]{E.A.~Kozyrev}
\author[adr1,adr2]{P.P.~Krokovny}
\author[adr1,adr2]{A.E.~Kuzmenko}
\author[adr1,adr2]{A.S.~Kuzmin}
\author[adr1,adr2]{I.B.~Logashenko}
\author[adr1,adr2]{P.A.~Lukin}
\author[adr1]{K.Yu.~Mikhailov}
\author[adr1]{V.S.~Okhapkin}
\author[adr1]{A.V.~Otboev}
\author[adr1]{Yu.N.~Pestov}
\author[adr1,adr2]{A.S.~Popov}
\author[adr1,adr2]{G.P.~Razuvaev}
\author[adr1]{Yu.A.~Rogovsky}
\author[adr1]{A.A.~Ruban}
\author[adr1]{N.M.~Ryskulov}
\author[adr1,adr2]{A.E.~Ryzhenenkov}
\author[adr1]{A.I.~Senchenko}
\author[adr1]{Yu.M.~Shatunov}
\author[adr1]{P.Yu.~Shatunov}
\author[adr1,adr2]{V.E.~Shebalin}
\author[adr1,adr2]{D.N.~Shemyakin}
\author[adr1,adr2]{B.A.~Shwartz}
\author[adr1,adr2]{D.B.~Shwartz}
\author[adr1,adr4]{A.L.~Sibidanov}
\author[adr1,adr2]{E.P.~Solodov\fnref{tnot}}
\author[adr1]{V.M.~Titov}
\author[adr1,adr2]{A.A.~Talyshev}
\author[adr1]{A.I.~Vorobiov}
\author[adr1]{I.M.~Zemlyansky}
\author[adr1,adr2]{Yu.V.~Yudin}

\address[adr1]{Budker Institute of Nuclear Physics, SB RAS, 
Novosibirsk, 630090, Russia}
\address[adr2]{Novosibirsk State University, Novosibirsk, 630090, Russia}
\address[adr3]{Novosibirsk State Technical University, 
Novosibirsk, 630092, Russia}
\address[adr4]{University of Victoria, Victoria, BC, Canada V8W 3P6}
\fntext[tnot]{Corresponding author: solodov@inp.nsk.su}


%
\vspace{0.7cm}
\begin{abstract}
\hspace*{\parindent}
The cross section of the process $e^+e^- \to \pipi\piz\eta$ has been measured
using a data sample of 21.8 pb$^{-1}$ collected with the CMD-3
detector at the VEPP-2000  $\epem$ collider.  2769$\pm$95 signal events 
have been selected in the center-of-mass energy range 1394--2005 MeV.
The production dynamics is dominated by the $\omega(782)\eta$ and 
$\phi(1020)\eta$ intermediate states in the lower energy range, 
and by the $a_0(980)\rho(770)$ intermediate state at higher energies.
\end{abstract}

\maketitle
\baselineskip=17pt
\section{ \boldmath Introduction}
\hspace*{\parindent}
The production dynamics of the $\pipi\piz\eta$ final state in $\epem$ 
annihilation has been never studied before. Only the $e^+e^-\to\omega(782)\eta$
cross section was measured by the BaBar Collaboration~\cite{isr6pi} with 
a relatively low statistical accuracy using $\eta\to\pipi\piz$ decay, and 
by the SND Collaboration~\cite{snd} (with $\eta\to\gamma\gamma$ decay).
The $e^+e^- \to \pipi\piz\eta$ cross section contributes a not negligble 
value (up to 15\% of the total hadronic cross section in some energy range) 
to the calculations of the hadronic contribution to the muon anomalous magnetic 
moment~\cite{g-2}, and a detailed study of the production dynamics can
further improve the accuracy of these calculations as well as our 
understanding of the spectroscopy of light mesons.

In this paper we report the analysis of the data sample based on 
21.8 pb$^{-1}$ of the integrated luminosity collected at the CMD-3 detector
in the 1394--2005 MeV center-of-mass energy (\Ecm) range. We identify the 
$\pipi\piz\eta$ candidate events using $\eta\to\gamma\gamma$ decay, and 
observe no candidate events below \Ecm= 1400 MeV. 
These data have been collected in three energy scans at 40 c.m. energy
points, performed  at  the VEPP-2000 collider~\cite{vepp} in 2011 and 2012.

The general-purpose detector CMD-3 has been described in 
detail elsewhere~\cite{sndcmd3}. Its tracking system consists of a 
cylindrical drift chamber (DC)~\cite{dc} and double-layer multiwire 
proportional Z-chamber, both also used for a trigger, and both inside a thin 
(0.2~X$_0$) superconducting solenoid with a field of 1.3~T.
The tracking system allows to detect charged tracks with a minimum polar 
angle about 0.5 radians relative to the beam axis (about 90\% of 4$\pi$).
The barrel liquid-xenon (LXe) calorimeter with a 5.4~X$_0$ thickness has
fine electrode structure, providing a 1--2 mm spatial resolution 
for photons~\cite{lxe}, and
shares the cryostat vacuum volume with the superconducting solenoid.     
The barrel CsI-crystal calorimeter is placed outside  the LXe calorimeter, 
and  increases the total thickness to   13.5~X$_0$.  The endcap BGO 
calorimeter with a thickness of 13.4~X$_0$ is placed inside the 
solenoid~\cite{cal}.
Our combined calorimeter allows to detect photons with a minimum polar 
angle down to 0.25 radians relative to the beam axis (about  98\% of 4$\pi$).
The luminosity is measured using events of Bhabha scattering 
at large angles with about 1\% accuracy~\cite{lum}. 

The beam energy has been monitored by measuring the current 
in the dipole magnets of the main ring, and at a few energy points by using 
the Back-Scattering-Laser-Light system~\cite{laser}. Using measured average 
momentum of Bhabha events, and average momentum of proton-antiproton pairs 
from the process $\epem\to p\bar p$~\cite{pbarp}, we determine \Ecm~at 
each energy point with about 1 MeV accuracy.  

To understand the detector response to processes under study and to
obtain a detection efficiency, we have developed Monte Carlo (MC) 
simulation of our detector based on the GEANT4~\cite{geant4} package, 
in which all simulated events pass the whole reconstruction and selection 
procedure. The MC simulation uses primary generators with matrix elements 
for the studied processes, including soft photon radiation by initial 
electron or positron, calculated according to Ref.~\cite{kur_fad}.
For the background study we have developed a special MC generator to simulate generically $\epem\to hadrons$, which includes the majority ($>$30) of exclusive channels weighted with their known cross sections. and perform analysis of events based on it. 
\section{Selection of $e^+e^-\to \pi^+\pi^-\pi^0\eta$ events}
\label{select}
\hspace*{\parindent}
Candidates for the process under study are required to have 
two good tracks of  charged particles with opposite charges, and four or more  
clusters in the calorimeters, not related to the tracks, assumed to be photons. 
We require ionization losses of a track in the DC to be consistent with the 
pion hypothesis, a  track momentum larger than 40~\mevc, a minimum 
distance from a track to the beam axis in the
transverse  plane less than 0.25 cm, and a minimum distance from a track 
to the center of the interaction region along the beam axis Z  less 
than 12 cm.
\begin{figure}[p]
\begin{center}
\vspace{-0.5cm}
\includegraphics[width=1.0\textwidth]{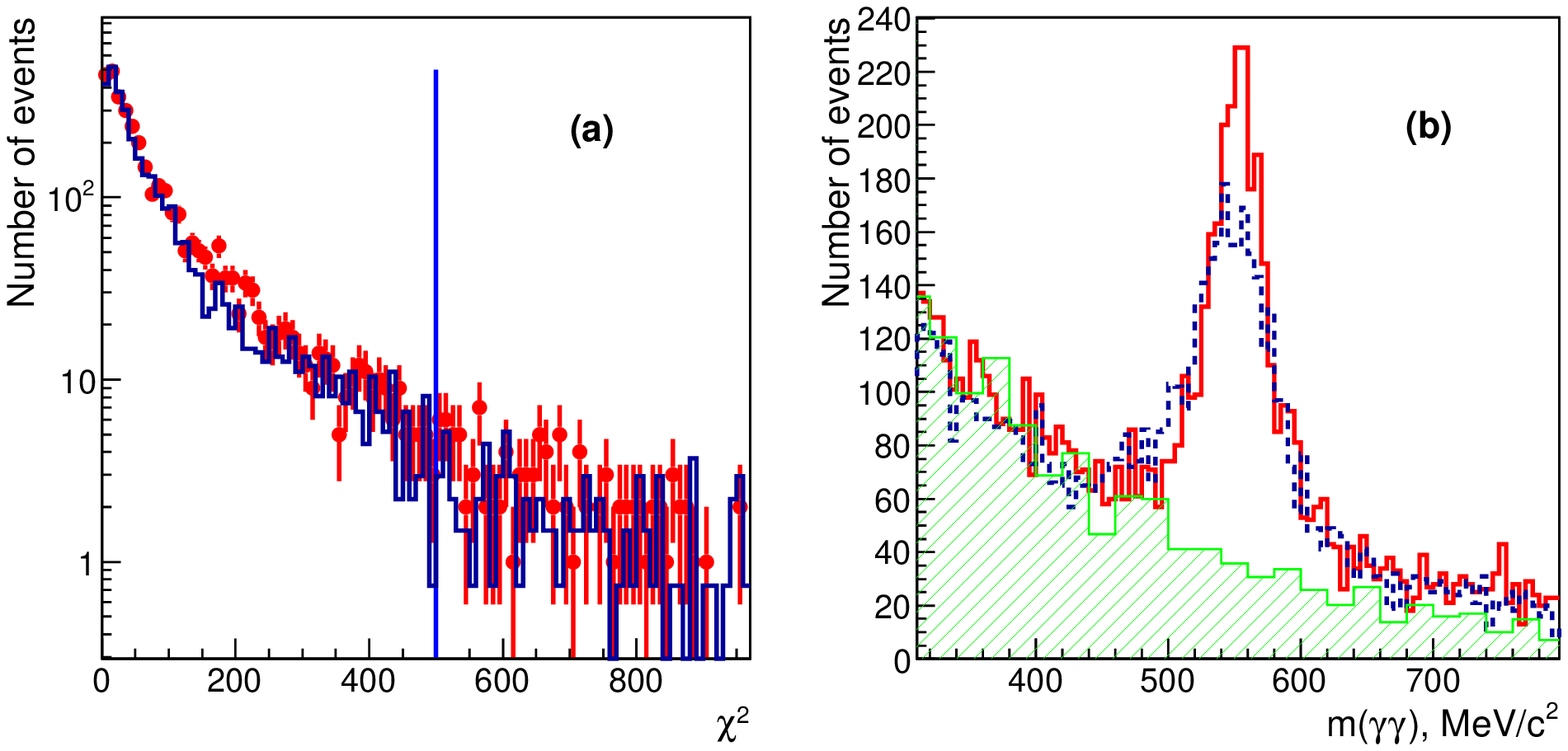}
\vspace{-0.7cm}
\caption
{(a)
The 5C-fit $\chi^2$ distribution for events with two tracks, \piz,~ and 
two photons for the $\epem\to\pipi\piz\gamma\gamma$ hypothesis 
for data (dots) and corresponding simulation (histograms).
(b) The experimental two-photon invariant mass distributions before (dashed histogram) 
and after (solid histogram) a kinematic fit. A shaded histogram is for the generic $\epem\to hadrons$ MC simulation with excluded signal process.
}
\label{chi2}
\end{center}
%
\begin{center}
\vspace{-0.5cm}
\includegraphics[width=1.0\textwidth]{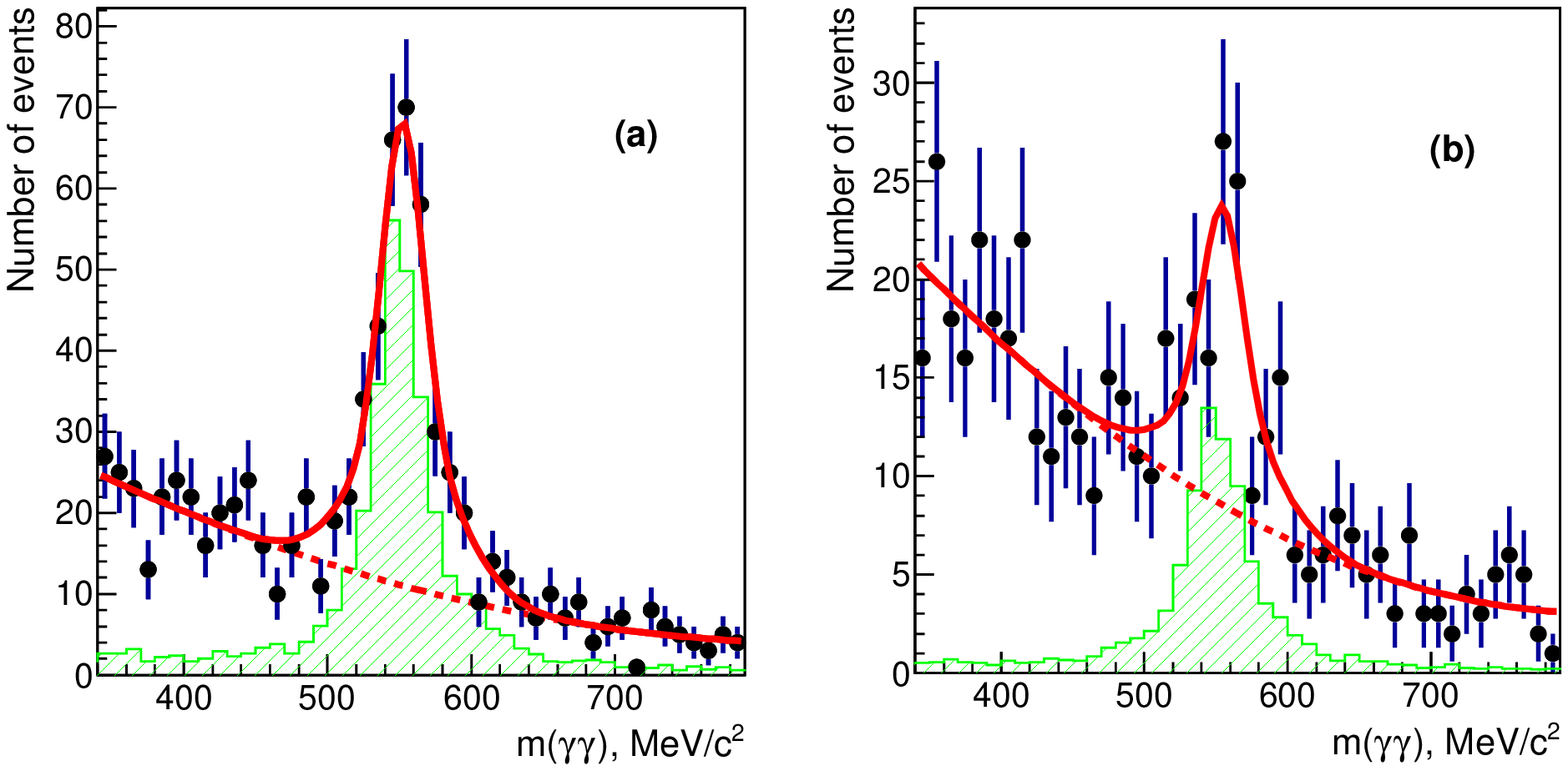}
\vspace{-0.7cm}
\caption
{
Two-photon invariant mass distributions and fit functions to determine 
the number of $\pipi\piz\eta$ events at \Ecm = 1680 MeV (a) and 
\Ecm = 1600 MeV (b). Dashed curves show the background contribution. Histograms are for the expected signal events from simulation. 
}
\label{etafit}
\end{center}
\end{figure}
The photon candidate is required to have energy deposition in the calorimeters 
exceeding 25 MeV.   

Reconstructed momenta and angles of the detected charged tracks as well as 
energy and angles of four photons are subject to the kinematic fit for 
the $\epem\to\pipi\piz\gamma\gamma$ hypothesis, assuming that the total
energy is equal to \Ecm~ and total momentum is equal to zero. 
First, we look for one photon pair with the invariant mass closest to the 
\piz~mass inside the $\pm$55~\mevcc~ (about $\pm$3.5 standard deviations) 
window, and we use the \piz~mass as an additional 
fifth constraint in the fit (5C fit) for this photon pair.  
 No additional constraints are applied to the second photon pair. 
The covariance matrices for charged tracks and photons are used in the fit 
and provide a $\chi^2$ value for each event.
A large fraction of the event candidates has 
more than four photons: we test all possible combinations, and two photon 
pairs with the smallest $\chi^2$ value are retained for further analysis. 
As a result of the fit, 
we obtain improved values of the momenta, energies and angles 
for all particles.

Figure~\ref{chi2}(a) shows the obtained $\chi^2$ distributions for the
experimental (dots) and simulated $\epem\to\pipi\piz\eta$ (histogram)
events, when the invariant mass of the second photon pair is in the 
$\pm$50~\mevcc~window around the $\eta$ mass. A vertical line shows 
the applied selection.  

Each event is also subject to the 4C fit under a 
$\epem\to\pipi\gamma\gamma$ hypothesis: all photon pairs are tested to get 
the best $\chi^2$ value, and a requirement $\chi^2_{\pipi\gamma\gamma}>40$ 
suppresses the background from the processes $\epem\to\pipi\piz$ and 
$\epem\to\pipi\eta$ by a factor of 10--20 to a negligible level with a 
1.5\% loss of the signal events. To study the remaining background we analyse events from the generic $\epem\to hadrons$ MC generator with the excluded signal process.

Figure~\ref{chi2}(b) presents the invariant mass distributions for the 
second photon pair before (dashed histogram) and after (solid histogram) the 
5C kinematic fit for events in the \Ecm= 1600-1800~MeV energy range 
and applied $\chi^2$ selection. A signal from the $\eta\to\gamma\gamma$ decay 
is clearly seen, and an improvement in the resolution is obtained. The shaded histogram shows a background from other processes, dominated by the $\epem\to\pipi\piz\piz$ reaction with wrong-assigned photons. No peaking background is observed.

The $\eta$ peak in the invariant mass distribution of the second photon pair 
is used to obtain the inclusive number of the $\pipi\piz\eta$ events. We fit  
the distributions of Fig.~\ref{chi2}(b) at each energy with a sum of functions
to separate signal and background. The shape of the $\eta$ signal is taken 
from the MC simulation (shown by shaded histograms), while a second-order polynomial function is used for 
the background. Two examples of the fit are shown in Fig.~\ref{etafit} at
\Ecm = 1680 MeV (a) and \Ecm = 1600 MeV (b). The total number of events 
evaluated by this procedure is $2769\pm95$. We do not observe any signal 
events for \Ecm below 1400~MeV, and present our data starting 
from \Ecm=1394 MeV.
\begin{figure}[tbh]
\begin{center}
\vspace{-0.5cm}
\includegraphics[width=1.04\textwidth]{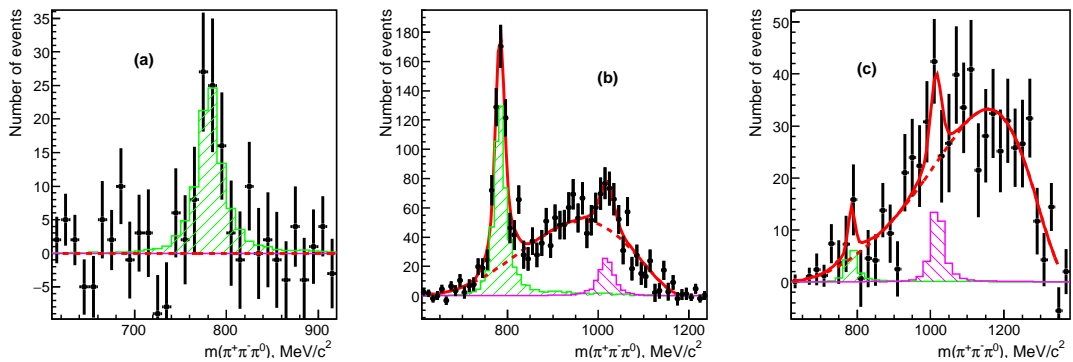}
\vspace{-0.3cm}
\caption
{
The background-subtracted $\pipi\piz$ invariant mass distribution for events 
in the \Ecm = 1400--1550 MeV (a), 1600--1800 MeV (b),  1800--2000 MeV (c) 
energy ranges. The lines show results of the fits with the $\omega(782)$ and $\phi(1020)$ 
signals (solid), and the background contribution (dashed). Histograms show 
expected MC-simulated signals from the
  $\omega(782)\eta$ and $\phi(1020)\eta$ intermediate states.
}
\label{3pimass}
\end{center}
\end{figure}

The observed $\pipi\piz\eta$ events contain several intermediate 
states. Our data sample is too small for standard amplitude analysis.
Instead, we first extract a contribution of the narrow intermediate resonances,
$\omega(782)$ and $\phi(1020)$, and then investigate other contributions, 
assuming low interference with the narrow states above.

\section{The $e^+e^-\to \omega(782)\eta$, $e^+e^-\to \phi(1020)\eta$ 
intermediate states}
\label{xsomegaeta}
\hspace*{\parindent}
To study intermediate states we select signal candidates by requiring
 $|m_{\gamma\gamma}-m_{\eta}| < 65$~\mevcc, see Fig.~\ref{chi2}(b), and subtract 
the sideband background using events with  
$65 < |m_{\gamma\gamma}-m_{\eta}| < 130$~\mevcc ~for any experimental distribution.
Figure~\ref{3pimass} shows the background-subtracted $\pipi\piz$ 
invariant mass distributions for the selected $\pipi\piz\eta$ candidates 
combined in three energy ranges: \Ecm = 1400--1550 MeV (a), 
\Ecm = 1600--1800 MeV (b), and \Ecm = 1800--2000 MeV (c). 
A signal from the $\omega(782)$ dominates at low energies, 
signals from the $\omega(782)$ and $\phi(1020)$ are well seen in the second 
range, and they are relatively small at higher energies, where other 
intermediate states dominate. 
To determine the number of $\omega$ and $\phi$ events, we fit distributions 
at each energy with a sum of the signal and combinatorial background functions. For the 
signal peaks we use double-Gaussian functions with all parameters, except the
number of events, fixed from the MC-simulation. A smooth function is used to 
describe the combinatorial background from other final states 
(see Sec.~\ref{xsa0rho}).  Histograms in Fig.~\ref{3pimass} show the 
expected MC-simulated signals from  the $\omega(782)\eta$ and 
$\phi(1020)\eta$ intermediate states. In total, for all energy points we obtain 
$824\pm41$ and $214\pm46$ events for the $\omega(782)\eta$ and 
$\phi(1020)\eta$ intermediate states, respectively. By variation of the 
polynomial order of the background function or removing sideband 
background subtraction, we estimate a systematic uncertainty on the number 
of signal events at about 5\%. 

\begin{figure}[p]
\begin{center}
\vspace{-0.3cm}
\includegraphics[width=1.\textwidth]{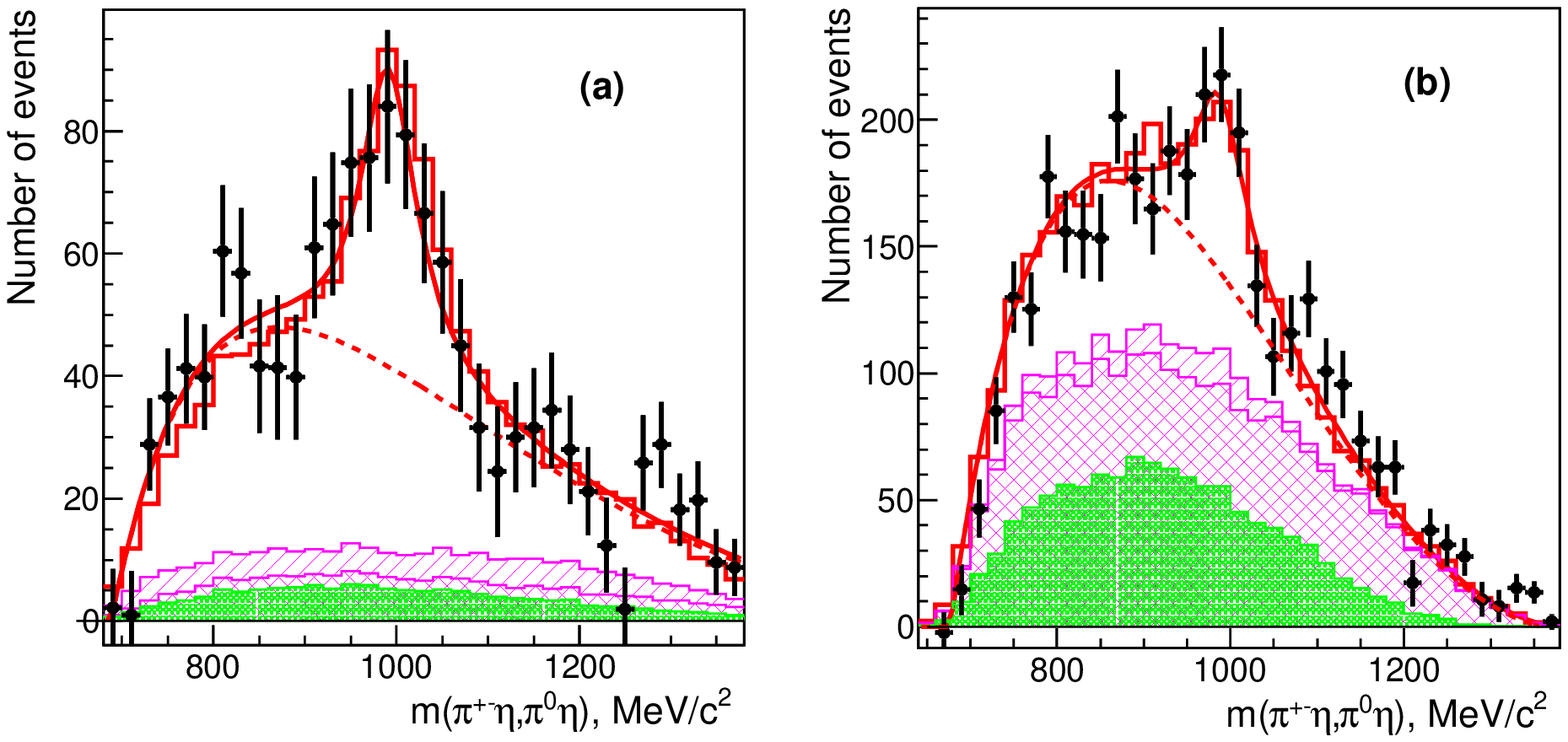}
\vspace{-0.4cm}
\caption{
The background-subtracted $\pi^+\eta, \pi^-\eta, \piz\eta$ invariant 
mass distributions for events in the 1800--2000 MeV (a) and 1600--1800 MeV 
(b) \Ecm~range from the $\eta$ peak of Fig.~\ref{chi2}(b). Histograms 
cumulatively show the MC-simulated contributions from the 
$\rho(1450,1700)\pi$ (shaded), $\omega(782)\eta$ (cross hatched), 
$\phi(1020)\eta$ (hatched), and $a_0(980)\rho(770)$ (open) intermediate states.
Lines show a fit with the $a_0(980)$ signal and background contribution 
(dashed). 
}
\label{mpieta}
\end{center}
%
%
\begin{center}
\vspace{-0.3cm}
\includegraphics[width=1.\textwidth]{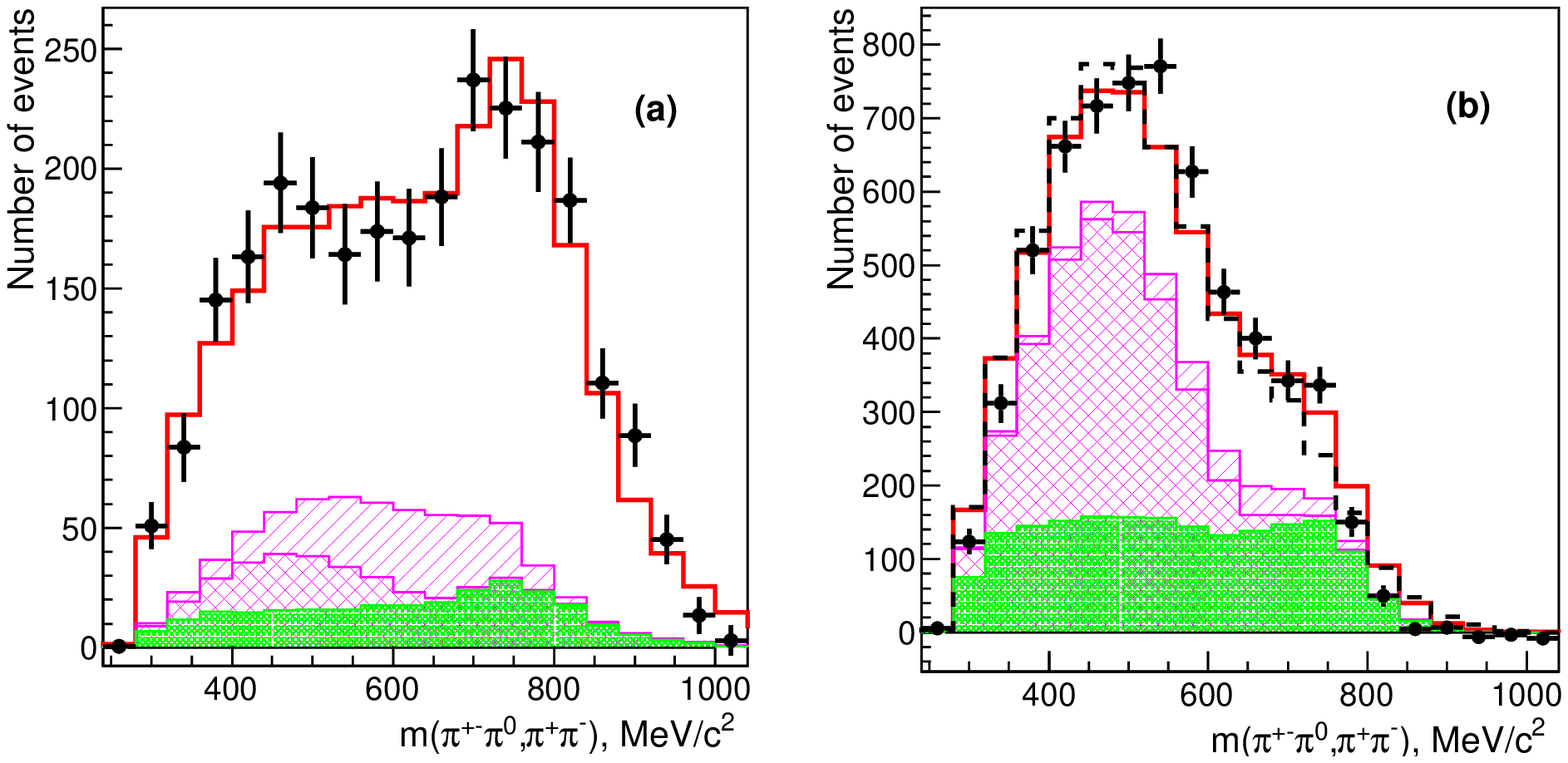}
\vspace{-0.4cm}
\caption{
The background-subtracted $\pi^+\piz, \pi^-\piz, \pipi$ invariant mass 
distributions for events in the 1800--2000 MeV (a) and 1600--1800 MeV (b) 
\Ecm ranges from the $\eta$ peak of Fig.~\ref{chi2}(b). Histograms 
cumulatively show the MC-simulated contributions from the 
$\rho(1450,1700)\pi$ (shaded), $\omega(782)\eta$ (cross hatched), 
$\phi(1020)\eta$ (hatched), and $a_0(980)\rho(770)$ (open) final states.
The dashed histogram in (b) corresponds to the case when 
the phase space $3\pi\eta$ simulation is used instead of the 
$\rho(1450,1700)\pi$ final state.
}
\label{rhoprime}
\end{center}
\end{figure}
\section{The $\epem\to a_0(980)\rho(770)$ intermediate state}
\label{xsa0rho}
\hspace*{\parindent}
The combinatorial background for the $\omega(782)\eta$ and $\phi(1020)\eta$ final states is relatively large and other intermediate resonances are thus 
expected: most probable are $a_0(980)\rho(770)$ and  $\rho(1450,1700)\pi$ which have $\pipi\piz\eta$ at the end of the decay chains. The $a_0(980)$ is relatively narrow and should be seen in the $\eta\pi$ invariant mass.
Figure~\ref{mpieta} shows the background-subtracted 
$\pi^+\eta,~\pi^-\eta,~\piz\eta$ invariant mass distribution 
(three entries per event)
for the events in the \Ecm= 1800--2000 MeV (a) and \Ecm= 1600--1800 MeV (b)  
ranges. 
A clear signal associated with $a_0(980)$ is seen in both energy ranges: 
while events in Fig.~\ref{mpieta}(b) are below the nominal
$a_0(980)\rho(770)$  threshold, $a_0(980)$ is still visible due to 
the large width of the $\rho(770)$.
Histograms cumulatively  show MC-simulated contributions
from the expected $\rho(1450,1700)\pi$ (shaded, see discussion in Sec.~\ref{xsnores}), 
$\omega(782)\eta$ (cross hatched), $\phi(1020)\eta$ (hatched), and 
$a_0(980)\rho(770)$ (open) intermediate states.
Lines show a fit with the $a_0(980)$ signal and the contribution of 
the combinatorial background (dashed). We tune background-description function with the simulation, and the ``Fermi'' step for the threshold behavior 
convolved with the third-order polynomial function gives the best result.

 We also observe a clear signal from the $\rho(770)$ in the 
$\pipi,~\pi^-\piz,~\pi^+\piz$ corresponding mass combinations, 
shown in Fig.~\ref{rhoprime}(a) for the \Ecm= 1800--2000 MeV range, 
where the $a_0(980)\rho(770)$ final state dominates.
We fit the distributions of Fig.~\ref{mpieta} at each energy with a sum 
of functions describing signal, combinatorial and other backgrounds,  
shown by the lines in  Fig.~\ref{mpieta}. The $a_0(980)$ signal is fitted 
with a Breit-Wigner function using 55 MeV width~\cite{pdg} 
convolved with the detector resolution ($\approx 50$ MeV). We obtain 
$1072\pm116$ events corresponding to the process $\epem\to a_0(980)\rho(770)$.
We vary the shape of the function used for the combinatorial background 
subtraction and estimate a systematic uncertainty on the number of signal 
events as about 15\%.

\begin{figure}[tbh]
\begin{center}
\vspace{-0.2cm}
\includegraphics[width=0.49\textwidth]{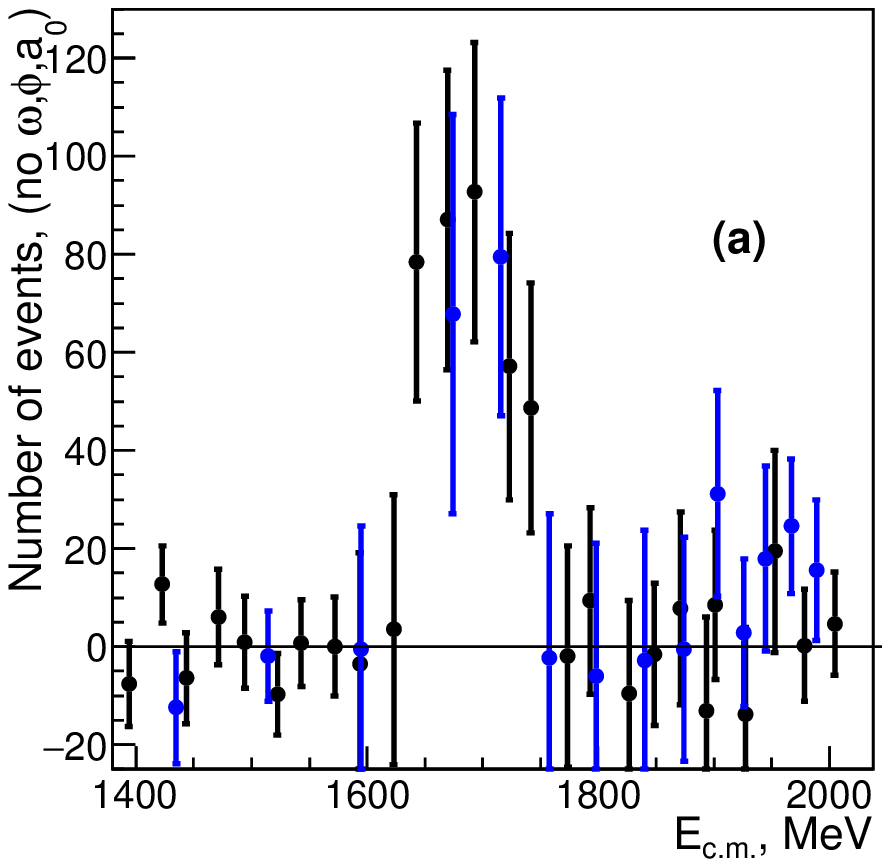}
\includegraphics[width=0.49\textwidth]{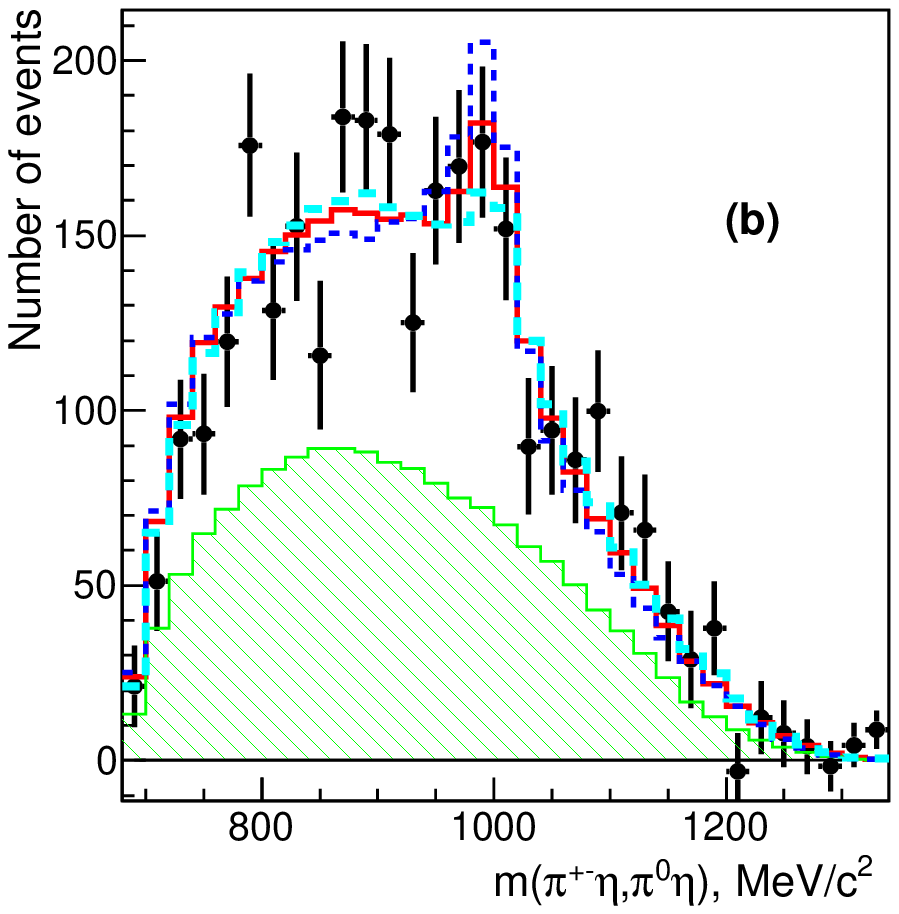}
\vspace{-0.3cm}
\caption
{
(a) The number of $\epem\to\pipi\piz\eta$ events with an excluded contribution 
from the $\omega(782)\eta$, $\phi(1020)\eta$, and $a_0(980)\rho(770)$ 
intermediate states. Black and blue colors are for the 2011 and 2012 data, 
respectively. 
(b) The background-subtracted $\pi^{+-0}\eta$ (three entries per event) 
invariant mass distribution with excluded  $\omega(782)\eta$ and 
$\phi(1020)\eta$ intermediate states. Short-dashed, solid, and long-dashed 
histograms shows a simulated signal from a sum of the $a_0(980)\rho(770)$ 
and  $\rho(1450,1700)\pi$ intermediate states in case of constructive, 
no-interference, and destructive interference of their amplitudes, 
respectively. The shaded histogram shows the contribution of the  
$\rho(1450,1700)\pi$ intermediate state only. 
}
\label{evn_noomphi}
\end{center}
\end{figure}
\section{The $\epem\to\pipi\piz\eta ({\bf{no}}~\omega, \phi, a_0)$ 
intermediate state}
\label{xsnores}
\hspace*{\parindent}
In the \Ecm= 1900--2000 MeV energy range the number of the 
$\pipi\piz\eta$ events
determined from the $\eta$ peak in Sec.~\ref{etafit}
is almost completely dominated by the $a_0(980)\rho(770)$ channel. 
The obtained number of $\pipi\piz\eta$ events in the \Ecm= 1600--1800 MeV 
energy range exceeded that expected from the sum of the 
$\epem\to\omega(782)\eta$ (33\%), 
$\phi(1020)\eta$ (7\%), and $a_0(980)\rho(770)$ (29\%) reactions. 

At each \Ecm~ energy we subtract events obtained for the $\omega, \phi, a_0$ 
signals from the total number of events obtained from the $\eta$ signal of 
Fig.~\ref{chi2}(b), and show the difference vs \Ecm~ in 
Fig.~\ref{evn_noomphi}(a). A resonant structure is observed around 
\Ecm~ = 1700 MeV.   
The difference (about 31\%) can be, for example, explained by the presence 
of the $\epem\to\omega(1650)\to\rho(1450,1700)\pi\to\rho(770)\eta\pi$ reaction.
In this case an additional signal from the $\rho(770)$ should be seen in 
two-pion masses. For the \Ecm= 1600--1800 MeV energy range we
show the background-subtracted  $\pi^+\piz, \pi^-\piz, \pipi$ invariant 
mass distribution  (three entries       per event) in Fig.~\ref{rhoprime}(b).
Histograms cumulatively show the MC-simulated contributions from the 
$\rho(1450,1700)\pi$
(shaded), $\omega(782)\eta$ (cross hatched), $\phi(1020)\eta$ (hatched), 
and $a_0(980)\rho(770)$ (open) final states. 
The dashed histogram in Fig.~\ref{rhoprime}(b) presents simulation when the 
phase-space model is used instead of $\rho(1450,1700)\pi$, indicating some 
data excess around the $\rho(770)$ mass.
Our data do not contradict to the presence of the 
$\epem\to\omega(1650)\to\rho(1450,1700)\pi\to\rho(770)\eta\pi$ reaction, 
but an additional $\rho(770)$ contribution  cannot be quantitatively 
extracted from the mass distributions with reasonable accuracy. Note 
that the $\rho(770)$ signal from the $a_0(980)\rho(770)$ intermediate state 
is also diluted below 1750~GeV due to limited phase space.

Moreover, the expected  $\rho(1450,1700)\pi\to\rho(770)\eta\pi$ decay chain 
and $a_0(980)\rho(770)$ both contain a relatively broad $\rho(770)$ resonance, 
and can interfere at the amplitude level. To examine an interference effect 
we sum two equal amplitudes of the above intermediate states at the primary 
generator level, and perform simulation with positive and negative relative 
signs.  Figure~\ref{evn_noomphi}(b) shows  
the background-subtracted experimental $\pi^{+-0}\eta$ (three entries per event)
invariant mass distribution (points) with the excluded contribution from the  
$\omega(782)\eta$ and $\phi(1020)\eta$ intermediate states 
(using MC-simulation). Short-dashed, solid, and long-dashed histograms show 
a simulated signal from  the sum of the $a_0(980)\rho(770)$ and  
$\rho(1450,1700)\pi$ intermediate states in case of constructive, 
no-interference, and destructive interference of the amplitudes, respectively. 
The shaded histogram shows a contribution of the  $\rho(1450,1700)\pi$ 
intermediate state only. When we fit the $a_0(980)$ signal peak as described above, the number of events changes by $\pm$50\% from the value with no interference. Because of that,
we should add at least a 50\% model-dependent systematic error to 
the number of $a_0(980)\rho(770)$ (and hence to $\rho(1450,1700)\pi$) 
events in the \Ecm~=1650--1750 MeV energy range, where overlap is the largest. 

\section{Detection efficiency}
\label{eff}
\begin{figure}[p]
\begin{center}
\vspace{-0.5cm}
\includegraphics[width=1.0\textwidth]{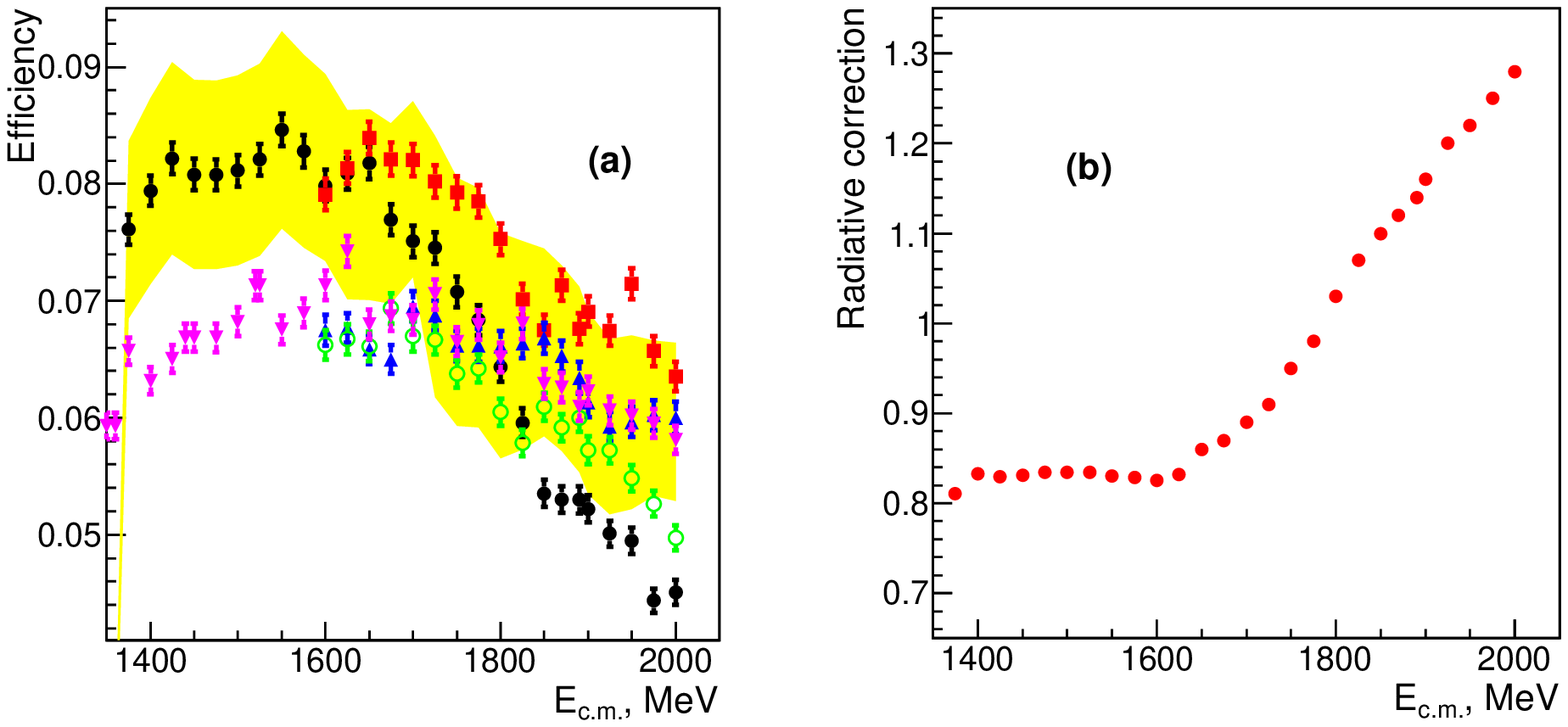}
\vspace{-0.7cm}
\caption
{
(a) MC-calculated efficiency for different intermediate states: 
$\omega(782)\eta$ (circles), $\phi(1020)\eta$ (squares), $a_0(980)\rho(770)$ 
(triangles up), $\rho(1450,1700)\pi$ (open circles), 
phase space (triangles down); the shaded area shows an
average efficiency with estimated systematic errors.
(b) Radiative corrections ($1+\delta_R$) for the 
$e^+e^-\to \pi^+\pi^-\pi^0\eta$ cross section.
}
\label{eff2pi4g}
\end{center}
%
\begin{center}
\vspace{-0.5cm}
\includegraphics[width=1.0\textwidth]{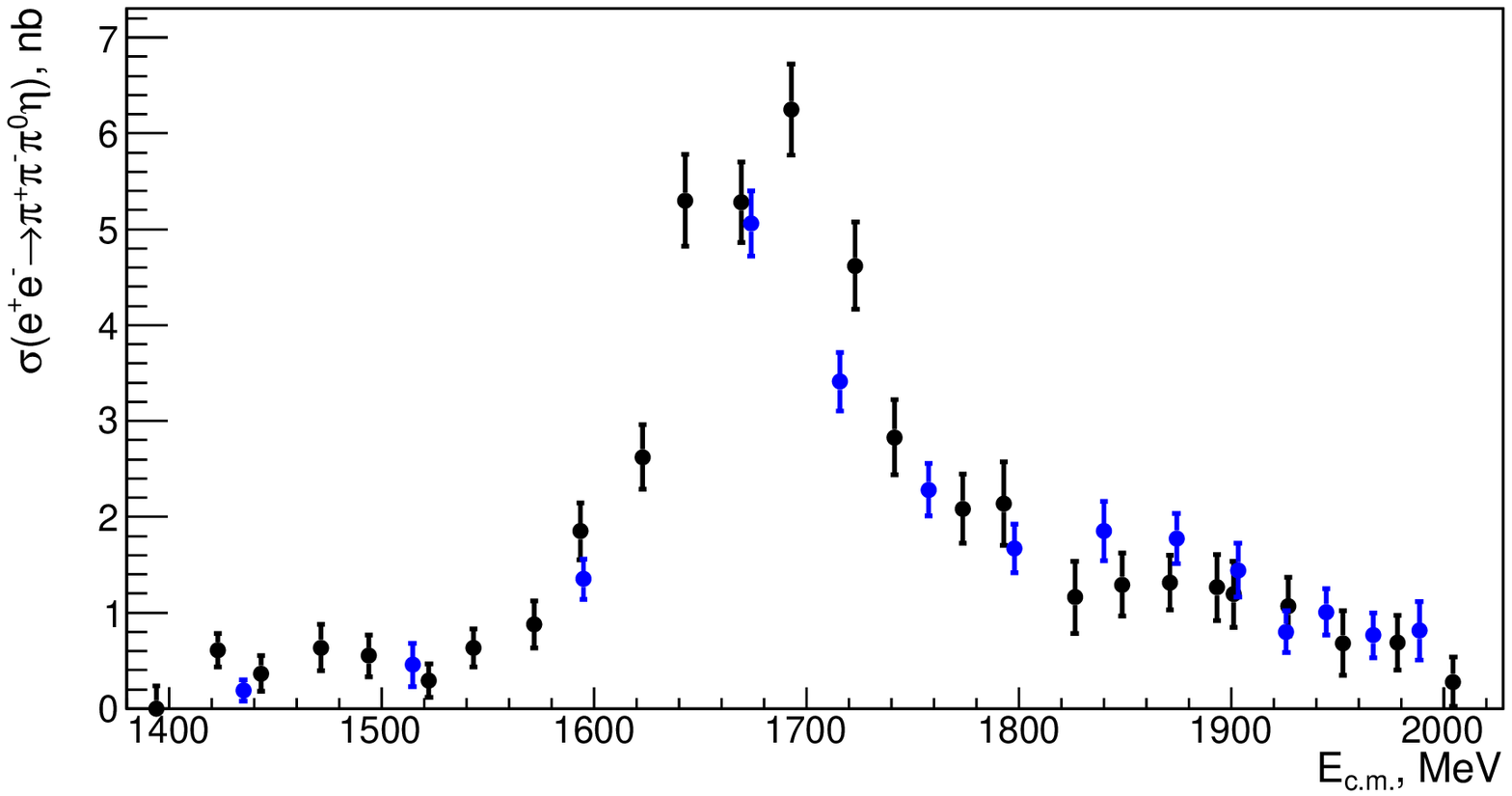}
\vspace{-0.7cm}
\caption
{
The $e^+e^-\to \pi^+\pi^-\pi^0\eta$ cross section obtained with the 
CMD-3 detector. Black and blue colors are for the 2011 and 2012 data, 
respectively. 
}
\label{xs_3pieta}
\end{center}
\end{figure}

As demonstarted above, the $\pipi\piz\eta$ final state is produced via 
several intermediate resonant states: we observe the $\omega(782)\eta$, 
$\phi(1020)\eta$, $a_0(980)\rho(770)$, and, possibly, 
$\rho(1450,1700)\pi\to\rho(770)\eta\pi$ intermediate states. Our detector 
does not have 100\% acceptance, and due to different angular 
distributions of final particles, we observe variations in the detection 
efficiency for different intermediate states. 
 Figure~\ref{eff2pi4g}(a) shows the MC-simulated 
$\epem\to\pipi\piz\eta$ detection efficiency ($\epsilon$) for different 
production modes determined as a ratio of events that passed reconstruction 
and selection criteria to the total number of simulated events.

As shown in Sec.~\ref{xsomegaeta}, the $\pipi\piz\eta$ final state  below 
\Ecm=1600 MeV is dominated by the process $\epem\to\omega(782)\eta$, 
an admixture of  the $\phi(1020)\eta$, $a_0(980)\rho(770)$,  and 
$\rho(1450,1700)\pi\to\rho(770)\eta\pi$ intermediate states arises in the 
\Ecm= 1600--1800 MeV range, and  the $a_0(980)\rho(770)$ state dominates 
above \Ecm=1800 MeV.

To estimate the detection efficiency for charged and neutral particles, we 
use a procedure with selecting a clean sample of events with one missing particle, predict momentum and angles of this particle using kinematics, and check how often this particle is detected with our detector. By applying this procedure to data and MC simulation we can obtain a correction for the calculated efficiency.
For this purpose we
use the $\epem\to\pipi\piz\piz$ process which has a much higher cross section 
in the studied energy range and low background. 

The correction to the MC-calculated efficiency of $-1.5\pm1.0$\% for a charged 
and $-1\pm1$\% for a neutral pion has been obtained.
Assuming similar efficiency for $\eta\to\gamma\gamma$ decay, we estimate 
the data-MC difference in the detection efficiency as a sum of corrections 
for two charged pions and two \piz's: $\epsilon_{\rm corr}=0.95$.   The uncertainty
of this number, $3$\%, obtained as a quadratic sum of 2\% from charged and 2\% 
from neutral pions, is taken as a systematic uncertainty. 

Our detection efficiency is obtained from MC simulation which includes a 
radiative photon from initial particles 
according to Ref.~\cite{kur_fad}, taking into account the energy dependence 
of each channel.

\begin{figure}[tbh]
\begin{center}
\vspace{-0.4cm}
\includegraphics[width=0.49\textwidth]{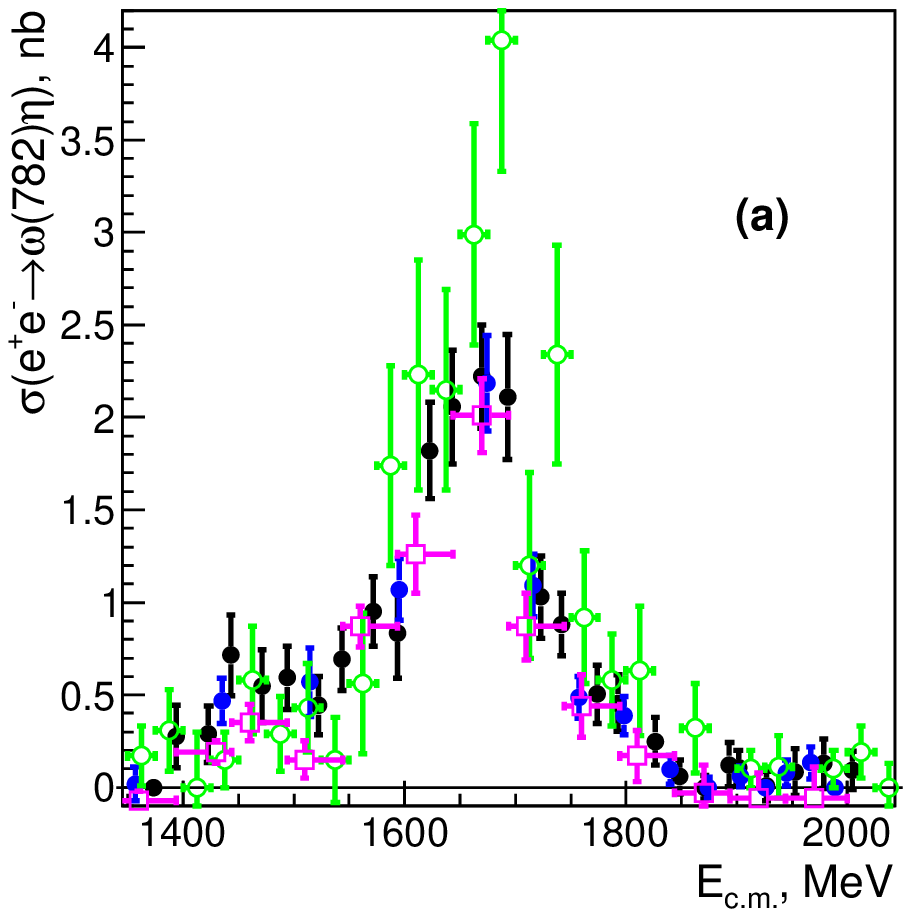}
\includegraphics[width=0.50\textwidth]{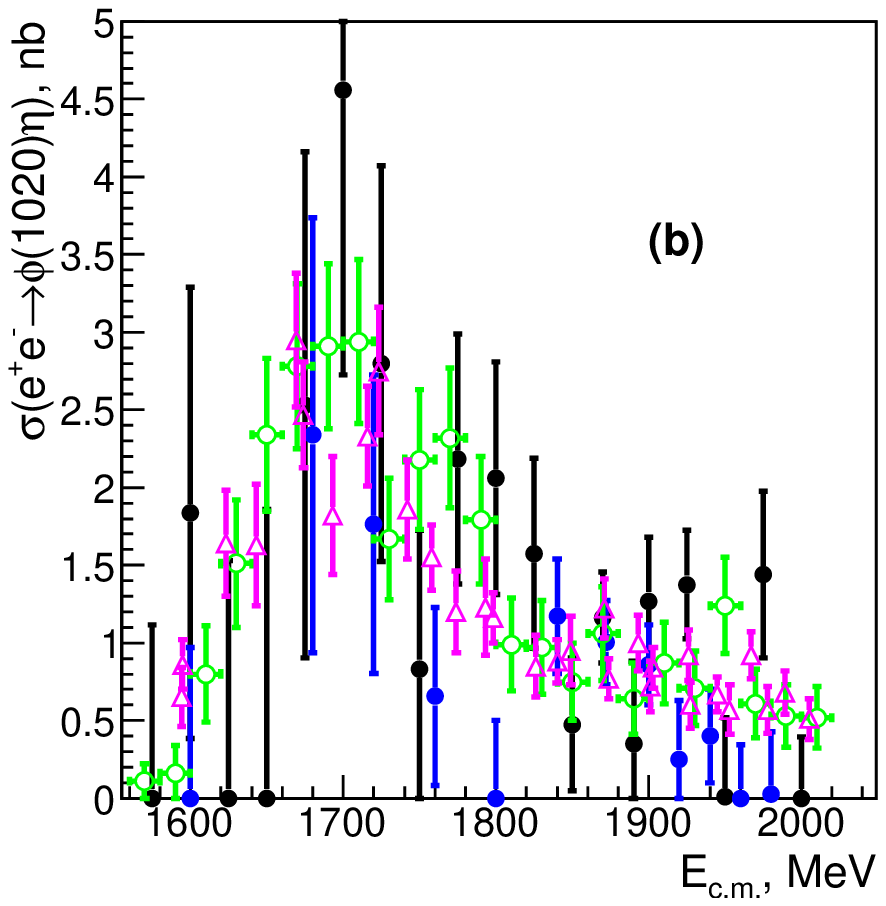}
\vspace{-0.3cm}
\caption
{
The $e^+e^-\to \omega(782)\eta$ (a) and  $e^+e^-\to \phi(1020)\eta$ (b) 
cross sections obtained at the CMD-3 detector in the 
$\pipi\piz\eta$ mode (circles). Black and blue colors are for the 2011 
and 2012 data, respectively. 
Also shown are corresponding measurements by BaBar (open circles), 
SND (open squares), and CMD-3 in the $K^+ K^-\eta$ mode (open triangles).
}
\label{xs_omphi}
\end{center}
\end{figure}
\section{The cross section calculation}
\label{xs2pipi0eta}

Using events of the process $e^+e^-\to \pi^+\pi^-\pi^0\eta$ or events 
of the intermediate states discussed above,
we calculate the cross sections at each energy as 
\begin{equation}
\label{xsformular}
\sigma(\pipi\piz\eta) = \frac{N}{L\cdot\epsilon\cdot 
(1+\delta_R)\cdot\epsilon_{\rm corr}}~,
\end{equation}
where $N$ is the number of selected events, $L$ is the integrated luminosity, 
$\epsilon$ is the detection efficiency shown in Fig.~\ref{eff2pi4g}(a) 
for all studied channels, and $(1+\delta_R)$ is a radiative correction. 
Since MC simulation does not perfectly describe the experimental 
resolutions, we apply a small correction, $\epsilon_{\rm corr}$, determined from 
the data as discussed in the Sec.~\ref{eff}.

To calculate the inclusive cross section for the process
$e^+e^-\to \pi^+\pi^-\pi^0\eta$, we use events obtained from the 
$\eta$ peak of Fig.~\ref{chi2}(b), and weight efficiencies calculated 
for different modes, taking into account the relative contribution of each 
channel. For this combined efficiency we
introduce a systematic uncertainty of about 10\%, shown as a shaded area in 
Fig.~\ref{eff2pi4g}(a).
The energy dependence of the  $(1+\delta_R)$ values is shown for this 
process in Fig.~\ref{eff2pi4g}(b): the values are obtained according to 
Ref.~\cite{kur_fad}, taking into account the energy dependence of the 
observed cross section (by iteration), presented in 
Fig.~\ref{xs_3pieta} and listed in Table~\ref{xs_all}. It is the first 
measurement of this cross section.

Using Eq.~\ref{xsformular} we calculate the cross sections for the 
processes $e^+e^-\to \omega(782)\eta$ and $e^+e^-\to \phi(1020)\eta$. 
Radiative corrections for the $\omega(782)\eta$ and $\phi(1020)\eta$ 
final states are calculated according to Ref.~\cite{kur_fad}. 
Efficiencies for these two processes are determined from the simulation  
(see discussion in Sec.~\ref{eff}) and shown in Fig.~\ref{eff2pi4g}(a).
Figure~\ref{xs_omphi} shows the obtained cross sections for the processes 
$e^+e^-\to \omega(782)\eta$ (a) and $e^+e^-\to \phi(1020)\eta$ (b) in 
comparison with previous measurements by BaBar~\cite{isr6pi}, SND~\cite{snd}, 
and CMD-3~\cite{phieta} for the $\phi(1020)\eta$ in the $K^+K^-\eta$ mode. 
The branching fractions of the $\omega(782)\to\pipi\piz$ and 
$\phi(1020)\to\pipi\piz$~decays\cite{pdg} are taken into account. Our data 
for the $\epem\to\omega(782)\eta$ cross section, listed in 
Table~\ref{xs_all}, are in good agreement with the SND experiment, 
and confirm a discrepancy with the BaBar data. 
The number of the $\omega(782)\eta$ events in the energy range below 
\Ecm= 1600 MeV is equivalent to the number of $\pipi\piz\eta$ events 
discussed in Sec.~\ref{select}. 
The cross section for the $\phi(1020)\eta$ mode is compatible with the 
previous measurements, but has much lower statistical precision because of 
a small $\phi\to\pipi\piz$ branching fraction.

\begin{figure}[tbh]
\begin{center}
\vspace{-0.2cm}
\includegraphics[width=0.50\textwidth]{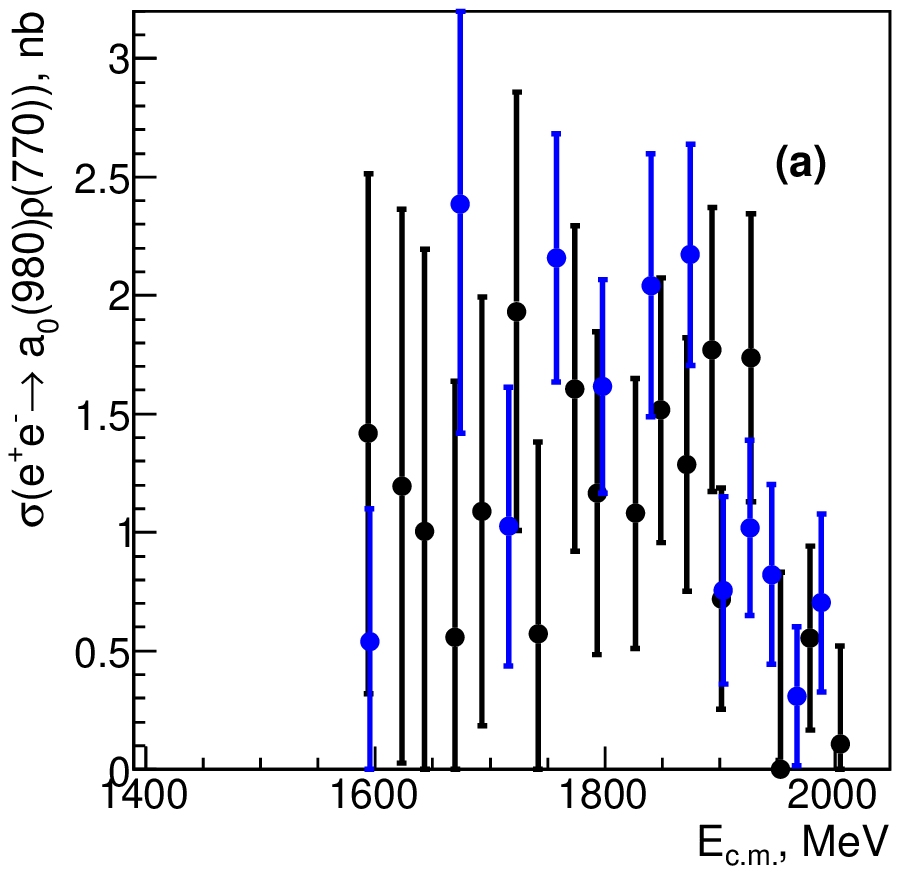}
\includegraphics[width=0.48\textwidth]{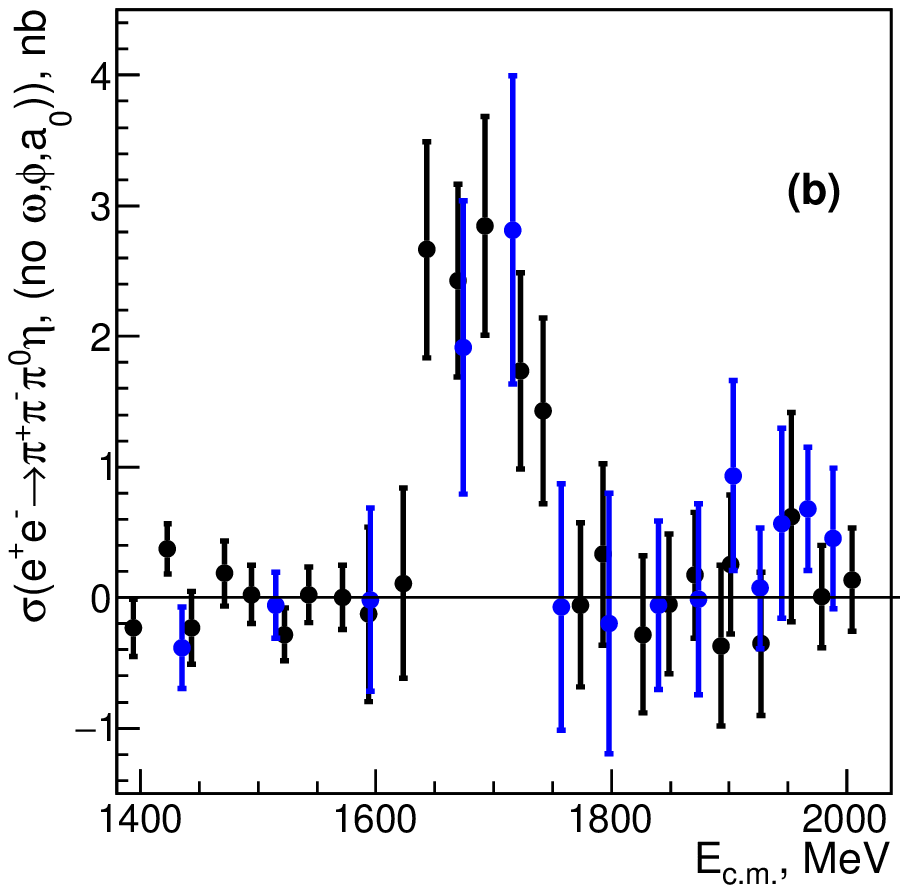}
\vspace{-0.4cm}
\caption{
(a) The $e^+e^-\to a_0(980)\rho(770)$  cross sections obtained at the 
\mbox{CMD-3} detector in the $\pipi\piz\eta$ mode.
(b) The cross section for the process
$\epem\to\pipi\piz\eta ({\bf{no}}~\omega, \phi, a_0)$, obtained at 
the CMD-3 detector.
}
\label{xsrhorhop}
\end{center}
\end{figure}

Figure~\ref{xsrhorhop}(a) shows the $\epem\to a_0(980)\rho(770)$ 
cross section calculated according to Eq.~\ref{xsformular} with 
efficiencies from Fig.~\ref{eff2pi4g}(b) (triangles up). It is the 
first measurement of this cross section, listed in Table~\ref{xs_all}.

Using the efficiency shown in Fig.~\ref{eff2pi4g}(a) (open circles) and 
the radiative correction similar to those shown in Fig.~\ref{eff2pi4g}(b), 
we calculate a cross section for the process 
$\epem\to\pipi\piz\eta ({\bf{no}}~\omega, \phi, a_0)$, presented in 
Fig.~\ref{xsrhorhop}(b) and listed in Table~\ref{xs_all}. The cross section 
has resonant behavior with a mass around \Ecm= 1700 MeV, but is consistent 
with zero below 1600 MeV and above 1800 MeV. The 
$\epem\to\omega(1650)\to\rho(1450,1700)\pi\to\rho(770)\eta\pi\to\pipi\piz\eta$ 
reaction chain can be responsible for this cross section.

\newpage
\section{Systematic errors and corrections}
\label{syst}
\hspace*{\parindent}

All cross sections above have a 1\% systematic uncertainty from the 
luminosity measurement~\cite{lum}, 3\% from inefficiency for charged and 
neutral pions (see Sec.~\ref{eff}), and 1\% from uncertainty on the 
radiative correction. 
Using two independent triggers based on DC or calorimeter information, 
the trigger efficiency is estimated to be close to unity with a 1\% systematic 
uncertainty. 

The uncertainties above should be combined with a ~5\% (15\%) uncertainty 
from variation of the signal and background shapes in the fitting procedure 
to extract $\omega(782)\eta$ ($\phi(1020)\eta$, $a_0(980)\rho(770)$) signals. 
We sum these errors in quadrature, and the~6.5\% (16\%) value is an overall 
systematic error for the measured exclusive cross sections. 

For the inclusive process $\epem\to\pipi\piz\eta$ we add a 10\% 
(11\% total) systematic uncertainty due to variation of efficiencies 
of contributing channels.

And finally, for the process 
$\epem\to\pipi\piz\eta ({\bf{no}}~\omega, \phi, a_0)$,
assuming $\epem\to\rho(1450,1700)\pi$ reaction, we estimate the 
uncertainty on the cross section as about 50\% due to possible interference 
with the process $\epem\to a_0(980)\rho(770)$ in the \Ecm~= 1650--1750 MeV 
energy range, where the latter is also determined with the same uncertainty.
\section{Fit to the $\epem\to\omega(782)\eta$ reaction}
\label{omegafit}
\begin{figure}[tbh]
\begin{center}
\vspace{-0.5cm}
\includegraphics[width=1.0\textwidth]{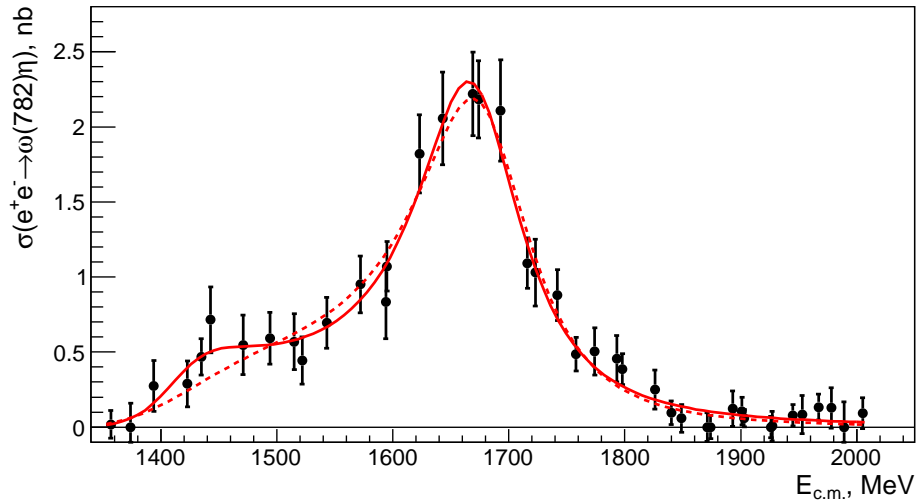}
\vspace{-0.7cm}
\caption
{
The $e^+e^-\to\omega(782)\eta$ cross sections obtained with the CMD-3 
detector in the $\pipi\piz\eta$ mode. Lines show Fit 1 (dashed line) and 
Fit 2 (solid line) described in the text.
}
\label{omfit_fig}
\end{center}
\end{figure}

\hspace*{\parindent} 
Using the procedure suggested in Ref.~\cite{snd,isr5pi}, we fit the  
$\epem\to\omega(782)\eta$ cross section with the sum of two 
$\omega$-like interfering resonances. The parameters of the $\omega(1420)$ 
(denoted below as $\omega'$) are not well determined~\cite{pdg}, and in our 
first fit we fix them at average values, similarly to Ref.~\cite{snd}.
A relative phase is fixed at $\pi$ to describe the asymmetry of the peak 
in the measured cross section.
Our results shown in Fig.~\ref{omfit_fig} by a dashed line and listed in 
Table~\ref{omfit_tab} (Fit 1), are consistent with that in 
Ref.~\cite{snd} (also shown in Table~\ref{omfit_tab}).  
The obtained width of the $\omega(1650)$ (denoted as $\omega''$) is 
significantly different from the values in PDG~\cite{pdg}, but close to 
that in Ref.~\cite{isr5pi} for the process $\epem\to\omega(782)\pipi$ (also 
shown in Table~\ref{omfit_tab}), and agrees with 
$\Gamma_{\omega''} = 114\pm14 $~MeV, obtained in Ref.~\cite{isr6pi}.
Our data allow us to perform a fit with floating  $\omega(1420)$ parameters, 
and the fit (Fit 2 in Table~\ref{omfit_tab} and the solid line in 
Fig.~\ref{omfit_fig})  yield the width smaller than estimated in PDG, 
but consistent with the result of Ref.~\cite{isr5pi}.

 In addition to the products of the $\omega', \omega''$ branching fractions 
to $\epem$ and the studied final state,  
$\BR_{ee}\BR_{\omega'f}, \BR_{ee}\BR_{\omega''f}$
in Table~\ref{omfit_tab}, we also calculate the products of electron width 
and branching fraction to final  state,
 $\Gamma_{ee}\BR_{\omega'f}, \Gamma_{ee}\BR_{\omega''f}$, which less depend on the 
uncertainty on the resonance widths. 
\begin{table*}[t]
\caption{
  Summary of parameters obtained from the fits described in the text.
  The values without errors were fixed in that fit.
  }
\label{omfit_tab}
\vspace{0.3cm}
\begin{tabular}{l c c c c c } 
\hline
Fit &  1 & 2 & $\omega\eta$~\cite{snd} & $\omega\pipi$~\cite{isr5pi} & PDG~\cite{pdg}\\
\hline
$\BR_{ee}\BR_{\omega'f}\cdot 10^7$    & 0.32$\pm$0.06     &  0.50$\pm$0.26
                                     &  $0.16^{+0.09}_{-0.07}$    
                                     &  1.3$\pm$0.4    &   --   \\
$\Gamma_{ee}\BR_{\omega'f}$(eV)       & 7.1$\pm$1.2      &  5.3$\pm$1.6
                                 &    -   
                                 &  17.5$\pm$5.4     &   --   \\
$m_{w'}$(\mevcc)          &   1420     &  1418$\pm$30  
                                  & 1420 
                                  & 1382$\pm$23       &  1400--1450 \\
$\Gamma_{\omega'}$(MeV)    &  220         & 104$\pm$35   
                                  & 220 
                                  & 133$\pm$48        &  180--250 \\
$\phi_{\omega'}$ (rad.)          &   $\pi$         &     $\pi$   
                                  &     $\pi$          
                                  &      $\pi$      &   -- \\
$\BR_{ee}\BR_{\omega''f}\cdot 10^7$    &  4.7$\pm$0.3    &  4.5$\pm$0.3
                                 &  4.4$\pm$0.5    
                                 &   4.7$\pm$0.4       &   --   \\
$\Gamma_{ee}\BR_{\omega''f}$(eV)   & 59$\pm$3   &  51$\pm$3
                                 &  -    
                                 &  103.5$\pm$8.3         &   --   \\
$m_{\omega''}$(\mevcc)          &  1679$\pm$5        &  1671$\pm$6   
                                  &  1660$\pm$10 
                                  &  1667$\pm$13       & 1670$\pm$30 \\
$\Gamma_{\omega''}$(MeV)    &  121$\pm$9         &  113$\pm$9   
                                  & 110$\pm$20 
                                  & 222$\pm$25        &  315$\pm$35 \\
$\chi^2$ /n.d.f.           &    23/35      &     18/33
                         &    14.5/9      
                         &    34.9/48      &     --     \\
\hline
\end{tabular}
\end{table*}                           

\begin{table}[tbh]
\caption{Number of signal events and the $\epem\to\pipi\piz\eta$,~$\omega(782)\eta$,~$a_0(980)\rho(770)$, and  $\pipi\piz\eta ({\rm{no}}~\omega,\phi, a_0)$ cross sections vs \Ecm, measured with the CMD-3 detector. Only statistical errors are shown.
}
\label{xs_all}
\vspace{-0.7cm}
\begin{center}
\renewcommand{\arraystretch}{0.85}
\begin{tabular}{cccccc}
\hline
{\Ecm, MeV}
&{$N(\pipi\piz\eta)$}
&{$\sigma(\pipi\piz\eta)$, nb}
&{$\sigma(\omega\eta)$, nb}
&{$\sigma(a_0\rho)$, nb}
&{$\sigma(\rm{no}~\omega,\phi, a_0)$, nb}\\ 
\hline
2005 & 10  $\pm$ 7  & 0.27 $\pm$ 0.25 & 0.09 $\pm$ 0.10 & 0.10 $\pm$ 0.41 & 0.13 $\pm$ 0.39 \\ 
1989 & 41  $\pm$ 12 & 0.81 $\pm$ 0.30 & 0.00 $\pm$ 0.16 & 0.70 $\pm$ 0.37 & 0.45 $\pm$ 0.53 \\ 
1978 & 25  $\pm$ 8  & 0.68 $\pm$ 0.28 & 0.12 $\pm$ 0.13 & 0.55 $\pm$ 0.38 & 0.00 $\pm$ 0.39 \\ 
1967 & 44  $\pm$ 11 & 0.76 $\pm$ 0.23 & 0.13 $\pm$ 0.08 & 0.30 $\pm$ 0.29 & 0.67 $\pm$ 0.47 \\ 
1953 & 22  $\pm$ 9  & 0.68 $\pm$ 0.33 & 0.08 $\pm$ 0.12 & 0.40 $\pm$ 0.83 & 0.61 $\pm$ 0.79 \\ 
1945 & 63  $\pm$ 12 & 1.00 $\pm$ 0.24 & 0.07 $\pm$ 0.07 & 0.82 $\pm$ 0.37 & 0.56 $\pm$ 0.72 \\ 
1927 & 42  $\pm$ 10 & 1.06 $\pm$ 0.30 & 0.00 $\pm$ 0.09 & 1.73 $\pm$ 0.60 & -0.35 $\pm$ 0.54 \\ 
1926 & 44  $\pm$ 10 & 0.80 $\pm$ 0.22 & 0.00 $\pm$ 0.07 & 1.01 $\pm$ 0.36 & 0.07 $\pm$ 0.46 \\ 
1903 & 79  $\pm$ 13 & 1.44 $\pm$ 0.28 & 0.06 $\pm$ 0.05 & 0.75 $\pm$ 0.39 & 0.93 $\pm$ 0.72 \\ 
1901 & 40  $\pm$ 10 & 1.19 $\pm$ 0.34 & 0.10 $\pm$ 0.09 & 0.72 $\pm$ 0.46 & 0.25 $\pm$ 0.52 \\ 
1893 & 45  $\pm$ 11 & 1.26 $\pm$ 0.34 & 0.12 $\pm$ 0.11 & 1.77 $\pm$ 0.59 & -0.36 $\pm$ 0.61 \\ 
1874 & 108 $\pm$ 14 & 1.77 $\pm$ 0.26 & 0.00 $\pm$ 0.07 & 2.17 $\pm$ 0.46 & -0.01 $\pm$ 0.73 \\ 
1871 & 60  $\pm$ 12 & 1.31 $\pm$ 0.28 & 0.00 $\pm$ 0.09 & 1.28 $\pm$ 0.53 & 0.17 $\pm$ 0.48 \\ 
1849 & 39  $\pm$ 9  & 1.29 $\pm$ 0.32 & 0.05 $\pm$ 0.09 & 1.51 $\pm$ 0.55 & -0.05 $\pm$ 0.53 \\ 
1840 & 102 $\pm$ 15 & 1.85 $\pm$ 0.30 & 0.09 $\pm$ 0.07 & 2.04 $\pm$ 0.55 & -0.05 $\pm$ 0.64 \\ 
1826 & 39  $\pm$ 12 & 1.15 $\pm$ 0.37 & 0.24 $\pm$ 0.12 & 1.08 $\pm$ 0.57 & -0.28 $\pm$ 0.6 \\ 
1798 & 103 $\pm$ 15 & 1.67 $\pm$ 0.25 & 0.38 $\pm$ 0.10 & 1.61 $\pm$ 0.44 & -0.19 $\pm$ 0.99 \\ 
1793 & 60  $\pm$ 12 & 2.13 $\pm$ 0.43 & 0.45 $\pm$ 0.15 & 1.16 $\pm$ 0.68 & 0.33 $\pm$ 0.69 \\ 
1774 & 74  $\pm$ 13 & 2.08 $\pm$ 0.36 & 0.50 $\pm$ 0.15 & 1.60 $\pm$ 0.68 & -0.05 $\pm$ 0.62 \\ 
1758 & 139 $\pm$ 17 & 2.28 $\pm$ 0.27 & 0.48 $\pm$ 0.11 & 2.15 $\pm$ 0.52 & -0.07 $\pm$ 0.94 \\ 
1742 & 96  $\pm$ 14 & 2.82 $\pm$ 0.39 & 0.87 $\pm$ 0.16 & 0.57 $\pm$ 0.80 & 1.43 $\pm$ 0.71 \\ 
1723 & 152 $\pm$ 16 & 4.61 $\pm$ 0.45 & 1.02 $\pm$ 0.22 & 1.93 $\pm$ 0.92 & 1.73 $\pm$ 0.74 \\ 
1716 & 203 $\pm$ 20 & 3.40 $\pm$ 0.30 & 1.09 $\pm$ 0.16 & 1.02 $\pm$ 0.58 & 2.81 $\pm$ 1.18 \\ 
1693 & 204 $\pm$ 17 & 6.24 $\pm$ 0.47 & 2.10 $\pm$ 0.33 & 1.08 $\pm$ 0.90 & 2.84 $\pm$ 0.83 \\ 
1674 & 273 $\pm$ 21 & 5.05 $\pm$ 0.34 & 2.18 $\pm$ 0.25 & 2.38 $\pm$ 0.96 & 1.91 $\pm$ 1.12 \\ 
1669 & 189 $\pm$ 17 & 5.27 $\pm$ 0.41 & 2.22 $\pm$ 0.27 & 0.55 $\pm$ 1.08 & 2.42 $\pm$ 0.73 \\ 
1643 & 156 $\pm$ 16 & 5.30 $\pm$ 0.47 & 2.05 $\pm$ 0.30 & 1.00 $\pm$ 1.19 & 2.66 $\pm$ 0.82 \\ 
1623 & 82  $\pm$ 13 & 2.62 $\pm$ 0.33 & 1.82 $\pm$ 0.26 & 1.19 $\pm$ 1.16 & 0.11 $\pm$ 0.73 \\ 
1595 & 75  $\pm$ 14 & 1.34 $\pm$ 0.20 & 1.07 $\pm$ 0.16 & 0.54 $\pm$ 0.55 & -0.01 $\pm$ 0.70 \\ 
1594 & 52  $\pm$ 10 & 1.84 $\pm$ 0.29 & 0.83 $\pm$ 0.24 & 1.41 $\pm$ 1.09 & -0.12 $\pm$ 0.66 \\ 
1572 & 30  $\pm$ 10 & 0.87 $\pm$ 0.24 & 0.95 $\pm$ 0.18 & -               & 0.00 $\pm$ 0.24 \\ 
1543 & 22  $\pm$ 8  & 0.63 $\pm$ 0.19 & 0.69 $\pm$ 0.16 & -               & 0.02 $\pm$ 0.21 \\ 
1522 & 10  $\pm$ 7  & 0.29 $\pm$ 0.17 & 0.44 $\pm$ 0.15 & -               & -0.28 $\pm$ 0.20 \\ 
1515 & 15  $\pm$ 9  & 0.45 $\pm$ 0.22 & 0.56 $\pm$ 0.18 & -               & -0.05 $\pm$ 0.25 \\ 
1494 & 19  $\pm$ 9  & 0.54 $\pm$ 0.21 & 0.59 $\pm$ 0.17 & -               & 0.02 $\pm$ 0.22 \\ 
1471 & 21  $\pm$ 9  & 0.63 $\pm$ 0.24 & 0.54 $\pm$ 0.19 & -               & 0.18 $\pm$ 0.25 \\ 
1443 & 10  $\pm$ 6  & 0.36 $\pm$ 0.18 & 0.71 $\pm$ 0.21 & -               & -0.23 $\pm$ 0.28 \\ 
1435 & 12  $\pm$ 8  & 0.18 $\pm$ 0.11 & 0.46 $\pm$ 0.12 & -               & -0.38 $\pm$ 0.31 \\ 
1423 & 21  $\pm$ 7  & 0.61 $\pm$ 0.17 & 0.28 $\pm$ 0.15 & -               & 0.37 $\pm$ 0.19 \\ 
1394 & 0   $\pm$ 9  & 0.00 $\pm$ 0.23 & 0.27 $\pm$ 0.16 & -               & -0.23 $\pm$ 0.21 \\ 
\hline
\end{tabular}
\end{center}
\end{table}

\section*{Conclusion}
\hspace*{\parindent} 
We report the first measurement of the $\epem\to\pipi\piz\eta$ 
($\eta\to\gamma\gamma$) 
cross section with the CMD-3 detector at the VEPP-2000 collider. 
We also present the cross sections for the intermediate states 
$\omega(782)\eta, \phi(1020)\eta$, and $ a_0(980)\rho(770)$. 

The  process $\epem\to\omega(782)\eta$ is well described by the sum of 
the $\omega(1420)$ resonance and the resonance 
with $m = 1671\pm 4\pm 10$~\mevcc, $\Gamma = 113 \pm 9\pm 10$~MeV, 
which could be associated with the  
$\omega(1650)$, but has a width smaller than suggested by PDG~\cite{pdg}. 

We observe a contribution to the  process $\epem\to\pipi\piz\eta$  
from the process not associated with the $\omega(782), \phi(1020)$, or $
a_0(980)$ intermediate states, which can be explained by the reaction 
$\epem\to\omega(1650)\to\rho(1450,1700)\pi\to\rho(770)\eta\pi$. 

\section*{Acknowledgment}
We thank the VEPP-2000 personnel for excellent machine operation. Part of this work related to the photon reconstruction algorithm in the electromagnetic calorimeter is supported by the Russian Science Foundation (project \#14-50-00080). The work is partially supported by the Russian Foundation for Basic Research (grants
RFBR 14-02-00580-a, RFBR 15-02-05674-a, RFBR 16-02-00160-a).

\end{document}